\documentclass[12pt]{iopart}
\usepackage{iopams}
\usepackage[utf8x]{inputenx}
\usepackage{graphicx}
\usepackage{hyperref}

\begin{document}

\title[Interface states ...]{Interface states in two-dimensional electron systems with spin-orbital interaction}

\author{Aleksei\,A\,Sukhanov and Vladimir\,A\,Sablikov}
\address{V.A.~Kotel'nikov Institute of Radio Engineering and Electronics,
Russian Academy of Sciences, Fryazino, Moscow District, 141190, Russia}

\begin{abstract}
Interface states at a boundary between regions with different spin-orbit interactions (SOIs) in two-dimensional (2D) electron systems are investigated within the one-band effective mass method with generalized boundary conditions for envelope functions. We have found that the interface states unexpectedly exist even if the effective interface potential equals zero. Depending on the system parameters, the energy of these states can lie in either or both forbidden and conduction bands of bulk states. The interface states have chiral spin texture similar to that of the edge states in 2D topological insulators. However, their energy spectrum is more sensitive to the interfacial potential, the largest effect being produced by the spin-dependent component of the interfacial potential. We have also studied the size quantization of the interface states in a strip of 2D electron gas with SOI and found an unusual (non-monotonic) dependence of the quantization energy on the strip width.
\end{abstract}

\pacs{73.20.-r, 73.21.-b}
\maketitle

\section{Introduction}

Spin-orbit interaction (SOI) produces plenty of fascinating effects in solids which provide broad possibilities for spin current generation and spin manipulation~\cite{Dyakonov}. A noteworthy feature of these effects is that in many cases the SOI acting in the bulk of the sample gives rise to effects which manifest themselves near the boundaries and interfaces with other media. It is enough to mention the following phenomena: (i) spin Hall effect~\cite{Dyakonov2,Hirsch,Murakami, Wunderlich,Kato}, where the spin current produced by an electric current gives rise to the spin accumulation at side boundaries; (ii) anomalous Hall effect~\cite{Nagaosa}, where a transverse voltage is generated by electrical current in spin polarized medium in the absence of external magnetic field; (iii) equilibrium edge spin currents in two-dimensional (2D) systems~\cite{Sablikov1} and edge spin accumulation~\cite{Usaj,Sonin}. The importance of the boundary effects has motivated us to study electronic states appearing at the heterointerfaces in the presence of SOIs.

The interface states in the systems without SOIs have been widely investigated using different approaches such as the envelope-function method, tight-binding and first-principle calculations~\cite{Foreman,Ivchenko,Takhtamirov}. However, in the presence of SOIs the interface states are not well studied yet, though this issue attracts now growing interest stimulated by rapid progress in the studies of topological insulators. Topological insulators are considered to be a new state of solids with inverted conduction and valence bands and strong SOI.~\cite{Hasan} The electron spectrum of topological insulators is characterized by the presence of edge or surface gapless states lying in the energy gap of bulk states. An essential property of the topological states is their chiral spin texture, due to which these states are protected against the scattering and robust to the variation of system parameters and boundary conditions.

The present paper aims to study the interface states in 2D electron systems with heterogeneous SOI, such as a contact of 2D regions with the SOIs of different kind or strength. Specifically, we consider contacts of regions with the Rashba and Dresselhaus SOIs and contacts of 2D regions with the SOI and the normal 2D electron gas without SOI. Interface states are studied within envelope-function approach with using generalized boundary conditions and an effective interface potential. We restrict ourselves by one-band model which is commonly used for 2D electron gas with SOI, in contrast to the case of topological insulators where at least two bands are to be taken into account to describe the edge states. Nevertheless, we find that in this system the interface states exist which are similar to the edge states in 2D topological insulators as regards chiral spin texture, but they are more sensitive to boundary conditions.

Depending on the parameters (such as the ratio of effective masses in adjacent regions and the band-bottom offset at the interface) the energy of these states can lie either in the forbidden band or in the conduction band for bulk states or in both bands. An interesting result is that the interface states exist even if the interface potential equals zero. Having analyzed conditions under which the interface states exist, we conclude that in 2D systems with heterogeneous SOIs there are the interface states that appear due to the bulk properties.

The dependence of the interface-state spectra on the parameters of the effective interfacial potential is studied within a model, in which the potential contains two components: a spin-independent component and a component arising from the SOI at the interface. The first component is shown to produce an energy shift of the spin-split dispersion curves, while the SOI component considerably changes the dispersion-curve form and the spin polarization of the states.

We study also the interface states in a 2D strip of finite width and find that the spatial overlap between the interface states at opposite edges of the strip essentially affects the spectrum, in addition to the usual size quantization effect. As a consequence of this effect the interface-state spectrum splits into two bands, the bottom of the lower band changing non-monotonically with the strip width.

The paper is organized as follows. In Sec.~\ref{eqs} the statement of the problem and basic equations are presented. In Sec.~\ref{SOI/N} we consider in detail the interface states at the boundary between a region with the SOI and a normal electron gas without the SOI. Sec.~\ref{RSOI/DSOI} presents the spectra of interface state at the contact of regions with the SOIs of different kinds (Rashba are Dresselhaus SOIs). In Sec.~\ref{Strip} the interface states in a strip are studied. We end with conclusions.

\section{Approaches and basic equations}
\label{eqs}

Consider a 2D electron system with a sharp heterointerface between two uniform regions with different SOIs. In each region the Hamiltonian for the one-band envelope function $\Psi$ is
\begin{equation}
 H_{i}=\frac{p^{2}}{2m_i}+U_i+H_{R,D}^{(i)} \,,
\label{H_i}
\end{equation}
where $i=1,2$ is the index of adjacent regions, $m_i$ is effective mass, $U_i$ is a potential energy, $H_{R,D}^{(i)}$ is the Hamiltonian of the Rashba (R) and Dresselhaus (D) SOIs
\begin{equation}
 H_{R}^{(i)}\!=\!\frac{\alpha_i}{\hbar}(p_{y}\sigma_{x}\!-\!p_{x}\sigma_{y}),\quad H_{D}^{(i)}\!=\!\frac{\beta_i}{\hbar}(p_{y}\sigma_{y}\!-\!p_{x}\sigma_{x})\,,
\end{equation}
$\sigma_x$ and $\sigma_y$ are the Pauli matrices, and $\alpha_i$ and $\beta_i$ are SOI strengths.

Boundary conditions for the envelope spinor functions \mbox{\boldmath$\Psi$}$^{(i)}$ at the contact ($x=0$) are expressed via the transfer matrix used in the effective-mass method~\cite{Ando,Tokatly}
\begin{equation}
\left(
\begin{array}{c}
 \mbox{\boldmath$\Psi$}^{(2)}(+0)\\
 \partial_x\mbox{\boldmath$\Psi$}^{(2)}(+0)
\end{array}
\right)=
\left(
\begin{array}{cc}
 \mathbf T_{11} & \mathbf T_{12}\\
 \mathbf T_{21} & \mathbf T_{22}
\end{array}
\right)
\left(
\begin{array}{c}
 \mbox{\boldmath$\Psi$}^{(1)}(-0)\\
 \partial_x\mbox{\boldmath$\Psi$}^{(1)}(-0)
\end{array}
\right)\,,
\label{boundary_gen}
\end{equation}
where matrices $\mathbf T_{ij}$ are determined by the symmetry of the system and specific structure of the interface.

The general solution of the Schr\"odinger equation in $i$-th region reads 
\begin{equation}
 \mbox{\boldmath$\Psi$}^{(i)}_{k^{(i)}_x,k_y}=e^{ik_yy}\sum\limits_j^{1,2}\sum\limits_s^{\pm}A^{(i)}_{j,s} \mbox{\boldmath$\chi$}^{(i)}_{j,s}e^{ik^{(i)}_{j,s}x}\,,
\label{wave_func}
\end{equation}
where $k_y$ is the tangential wavevector, $k^{(i)}_{j,s}$ is the $x$ component of the wavevector, defined by a characteristic equation of the Hamiltonian~(\ref{H_i}), $s$ stands for the spin index, $j$ numbers the solutions of the characteristic equation, $\mbox{\boldmath$\chi$}^{(i)}_{j,s}$ is the spin function.

Generally there is a set of four wavevectors $k^{(i)}_{j,s}$. For the SOI region they were described in detail in Refs~\cite{Sablikov,Tkach,Sukhanov}. A short resume is as follows. The wavevectors $k^{(i)}_{j,s}$ are complex functions of the energy $E$ and the tangential momentum $k_y$. Two of them correspond to the states propagating or decreasing along the $x$ axis, other two relate to the states propagating or decreasing in the opposite direction. In the energy range $E<U_i-E_{so}$, all $k^{(i)}_{j,s}$ contain both real and imaginary parts which describe decaying and oscillating states, $E_{so}$ is the characteristic energy of the SOI: $E_{so}=m\alpha^2/(2\hbar^2)$ for the Rashba SOI and $E_{so}=m\beta^2/(2\hbar^2)$ for the Dresselhaus SOI. When $E>U_i-E_{so}$, the wavevectors $k^{(i)}_{j,s}$ are either purely real or purely imaginary depending on the relation between $E$ and $k_y$.

To clarify whether the interface states exist near the boundary $x=0$ one needs to find the solutions satisfying the boundary conditions (\ref{boundary_gen}) and vanishing at infinity ($x\to\pm\infty$). Dropping the terms, which do not vanish at infinity, in Eq.~(\ref{wave_func}) we arrive to a system of homogeneous equations from Eq.~(\ref{boundary_gen}). The zeros of its determinant give equations for the interface-state spectrum.

The results obtained in such a way are very cumbersome since the $4\times 4$ matrix $\mathbf T$ contains too many elements in spite of the restrictions imposed by the time reversal symmetry and the Hermitian character of the matrix being taken into account. To simplify the problem we use hereafter the following model Hamiltonian of the interface
\begin{equation}
 H_b= v_0 \delta(x) + \frac{\gamma}{\hbar}p_{y}\sigma_{z} \delta(x)\,.
\label{boundary_H}
\end{equation}
which is widely used to describe the Tamm-like surface states at heterointerfaces within the envelope-function approach.~\cite{Ivchenko,Takhtamirov,Foreman} 

This Hamiltonian arises naturally with using the $\mathbf {k\cdot p}$ approximation, when the crystal potential step the interface is treated perturbatively ~\cite{Takhtamirov} or is introduced phenomenologically ~\cite{Ivchenko}. Here the first term is a spin-independent effective potential at the interface. The second term is the spin-dependent potential caused by the SOI originated from the crystal potential gradient at the interface.~\cite{Vasko,Takhtamirov} The interface parameters $v_0$ and $\gamma$ are determined by the microstructure of the interface and therefore can not be expressed in terms of bulk parameters of the materials only. The values of these parameters in realistic systems vary over a wide range. For instance, the parameter $\gamma$ can be as high as 3~eV$\cdot$\AA~ for the contact Bi/Ag(111).~\cite{Ast}

In addition to the $\delta$-like terms in the interface Hamiltonian, a term proportional to $\delta'(x)$ also appears in the perturbation theory of sharp heterojunctions.~\cite{Takhtamirov} This term leads to an envelope-function discontinuity at the interface. We do not include this term supposing that the discontinuity is weak.

With the Hamiltonian (\ref{boundary_H}), the transfer matrix takes the form:
\begin{equation}
 \mathbf T_{11}=\mathbf I,\quad \mathbf T_{12}= 0,\quad  \mathbf T_{22}=\mu^{-1}\mathbf T_{11},
\label{T111222}
\end{equation}
\begin{equation}
 \mathbf T_{21}\!\!=\!\frac{m_2}{\hbar^2}\!
\left(
\begin{array}{cc}
 2(v_0\!+\!\gamma k_{y})& \alpha_2\!-\!\alpha_1\!-\!i(\beta_2\!-\!\beta_1\!)\\
 \alpha_1\!-\!\alpha_2\!+\!i(\beta_1\!-\!\beta_2\!)&2(v_0\!-\!\gamma k_y)
\end{array}
\right)
\label{T21}
\end{equation}
where $\mu=m_1/m_2$. Note that only the diagonal elements of $\mathbf T_{21}$ matrix come from the interface potential while the others are determined by the bulk characteristics.

Below we present results of the interface-state spectrum calculations for several systems: the contact of SOI region and normal (N) electron gas (SOI/N), the contact of Rashba SOI (RSOI) region and Dresselhaus SOI (DSOI) region (RSOI/DSOI), the strip of electron gas with the Rashba SOI bounded laterally by DSOI regions (DSOI/RSOI/DSOI structure).

\section{Interface states in SOI/N contact}
\label{SOI/N}

\begin{figure}
\centerline{\includegraphics[width=0.65\linewidth]{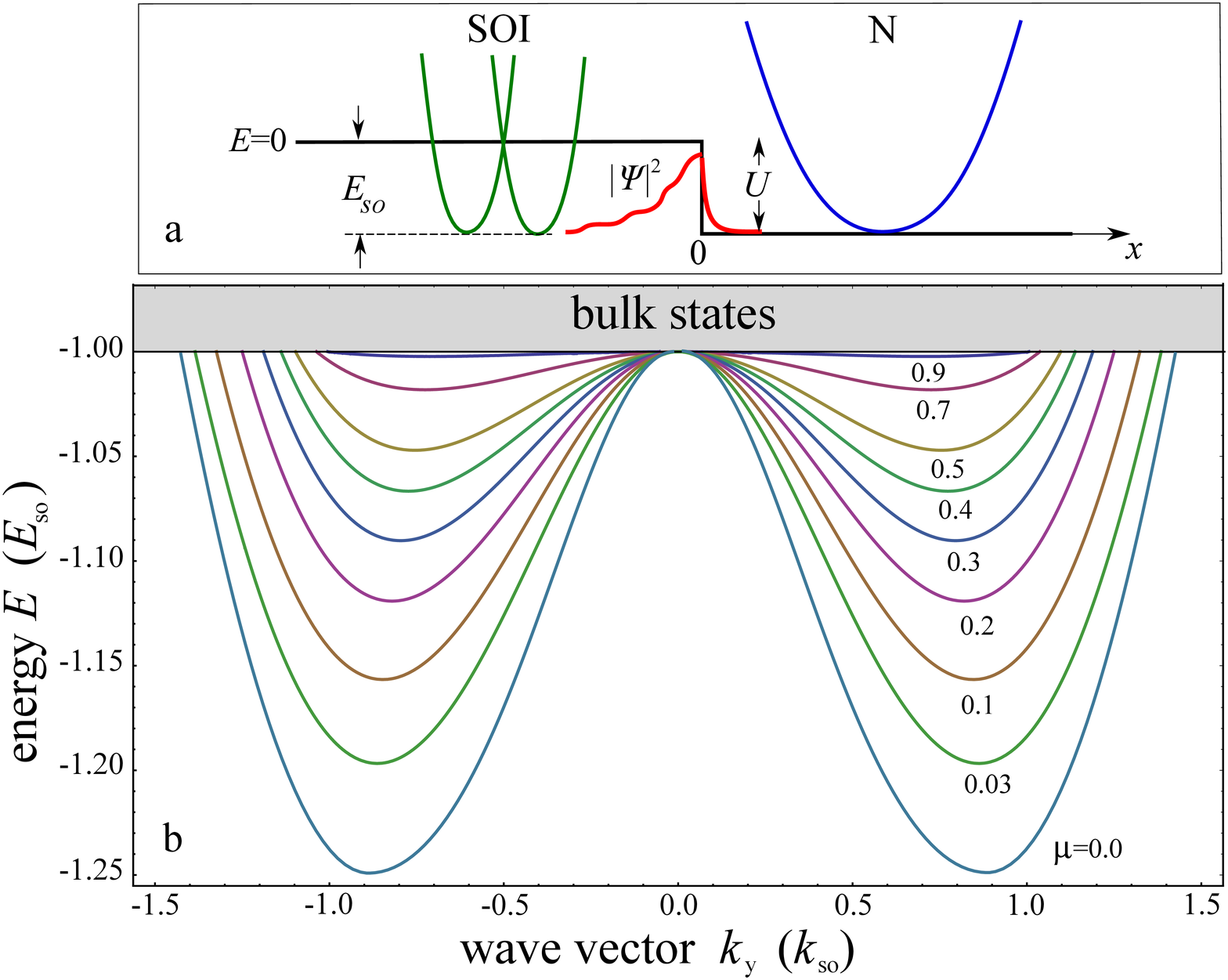}}
\caption{(Color online) (a) Energy diagram of the SOI/N contact, electron spectra in the bulk and the electron density distribution in an interface state. (b) Spectra of the interface states in the forbidden band for a variety of $\mu = 0.0\dots 0.9$ in the case where $U=E_{so}$.}
\label{SOI/N_f1}
\end{figure}

Consider a contact of the SOI region and normal 2D electron gas. To be specific suppose that the SOI is of Rashba type. In the case of Dresselhaus SOI the results are similar. The RSOI region is located at $x<0$ and N region lies at $x>0$. Let the potential energy in the SOI region be $U_{so}=0$ and in the N region be $U_N=-U$. The energy diagram is depicted in Fig~\ref{SOI/N_f1}a.

In the Rashba SOI region the wavefunction is
\begin{equation}
\mbox{\boldmath$\Psi$}^{(r)}=e^{ik_yy}\sum \limits_{s=\pm1}A_{s}
\left(
\begin{array}{c}
 \chi_{s}(\vec{k})\\
  1
\end{array}
\right) e^{\kappa_{s} x},
\label{eq2}
\end{equation}
where $\kappa_s=\kappa_1+s\kappa_2$,
\begin{equation}
\kappa_{1,2}=\frac{1}{\sqrt{2}}\sqrt{-\zeta\!+\!k_y^2-2k_{so}^{2} \pm \sqrt{\left(\zeta\!-\!k_y^2\right)^2\!-\!4k_{so}^2k_y^2}}\,,
\label{krx}
\end{equation}
\begin{equation}
\chi_s=\frac{2k_{so}(k_y+\kappa_{s})}{\zeta-k_y^2+\kappa_{s}^2}\,,
\label{chi_R}
\end{equation}
$\zeta=2m_{so}E/\hbar^2$, $k_{so}=m_{so}\alpha/\hbar^2$, $m_{so}$ is the effective mass of electrons in the SOI region.

In the N region the wavefunction is
\begin{equation}
\mbox{\boldmath$\Psi$}^{(N)}=e^{ik_yy}
\left[t_{1}
\left(
\begin{array}{c}
 1\\
 0
\end{array}
\right)+t_{2}
\left(
\begin{array}{c}
 0\\
 1
\end{array}
\right)
\right] e^{-g x}\,,
\label{wave-n}
\end{equation}
where
$g=\sqrt{k_{y}^{2}-(\zeta+u)/\mu}$, $u=2 m_{so}U/\hbar^2$, $\mu=m_{so}/m_N$, $m_N$ being the effective mass in the N region.

\subsection{Interface-state spectra}
\label{Spectra}
By matching the wavefunctions at $x=0$ with the use of Eqs~(\ref{boundary_gen}),(\ref{T111222}),(\ref{T21}) we come to the following condition under which the wavefunction amplitudes are nonzero:
\begin{equation}
\begin{array}{l}
 \left[(\kappa_1+\mu g+\bar{v}_0)^2-\kappa_2^2+k_{so}^2-(\bar{\gamma} k_y)^2\right](\chi_+-\chi_-)\\
 -2\bar{\gamma} k_y \kappa_2 (\chi_+ + \chi_-)+2 k_{so} \kappa_2 (1+\chi_+ \chi_-)=0 \,,
\end{array}
\label{SOI/N_1}
\end{equation}
where $\bar{v}_0=2v_0m_{so}/\hbar^2$ and $\bar{\gamma}=2\gamma m_{so}/\hbar^2$. In Eq.~(\ref{SOI/N_1}), $g$ and $\kappa_{1,2}$ are the functions of $\zeta$ and $k_y$ defined such that
\begin{equation}
\mathrm {Re}\,g(\zeta,k_y)>0\,,\quad   \mathrm {Re}\,\kappa_s(\zeta,k_y)>0\,.
\label{add cond}
\end{equation}
Eqs.~(\ref{add cond}) are additional conditions to Eq.~(\ref{SOI/N_1}).

Taking into account the explicit dependences of $\kappa_1$, $\kappa_2$, $g$ and $\chi_s$ on $\zeta$ and $k_y$ [given by Eqs~(\ref{krx}), (\ref{chi_R}) and (\ref{wave-n})] and Eq.~(\ref{SOI/N_1}) we arrive at the interface-state spectrum: $\zeta=\zeta_{ES}(k_y)$.

This equation is rather cumbersome in the full form. To analyze it we first consider a simple case where the interface Hamiltonian (\ref{boundary_H}) is absent. Assuming that $v_0=0$ and $\gamma=0$, Eq.~(\ref{SOI/N_1}) is simplified to
\begin{equation}
(\kappa_1+\mu g)^2-\kappa_2^2-k_{so}^2=0\,.
\label{eq11}
\end{equation}
This equation describes the interface states which appear due to the SOI in the bulk. The analysis of Eq.~(\ref{eq11}) shows that the interface states in the forbidden band exist only if $\mu<1$ ($m_{so}<m_N$) and the potential of the N region is higher than $-1.25E_{so}$. The interface-state spectra for a variety of values of $\mu$ and $U=E_{so}$ are presented in Fig.~\ref{SOI/N_f1}b. All dispersion curves lie above the curve
\begin{equation}
 E_0(k_y)=\frac{\hbar^2}{2m_{so}}\left( k_y^2-|k_{so}|\sqrt{k_{so}^2+4k_y^2}\,\right)\,,
\label{mu=0}
\end{equation}
that corresponds to the limiting case $\mu\to 0$. It is seen that the interface states exist below the conduction band bottom in the SOI region in the energy interval $-1.25E_{so}^2\leq E \leq -E_{so}^2$. At a given energy there are two pairs of the interface states with different signs of the wavevector $k_y$ and the group velocity.

\begin{figure}
\centerline{\includegraphics[width=0.55\linewidth]{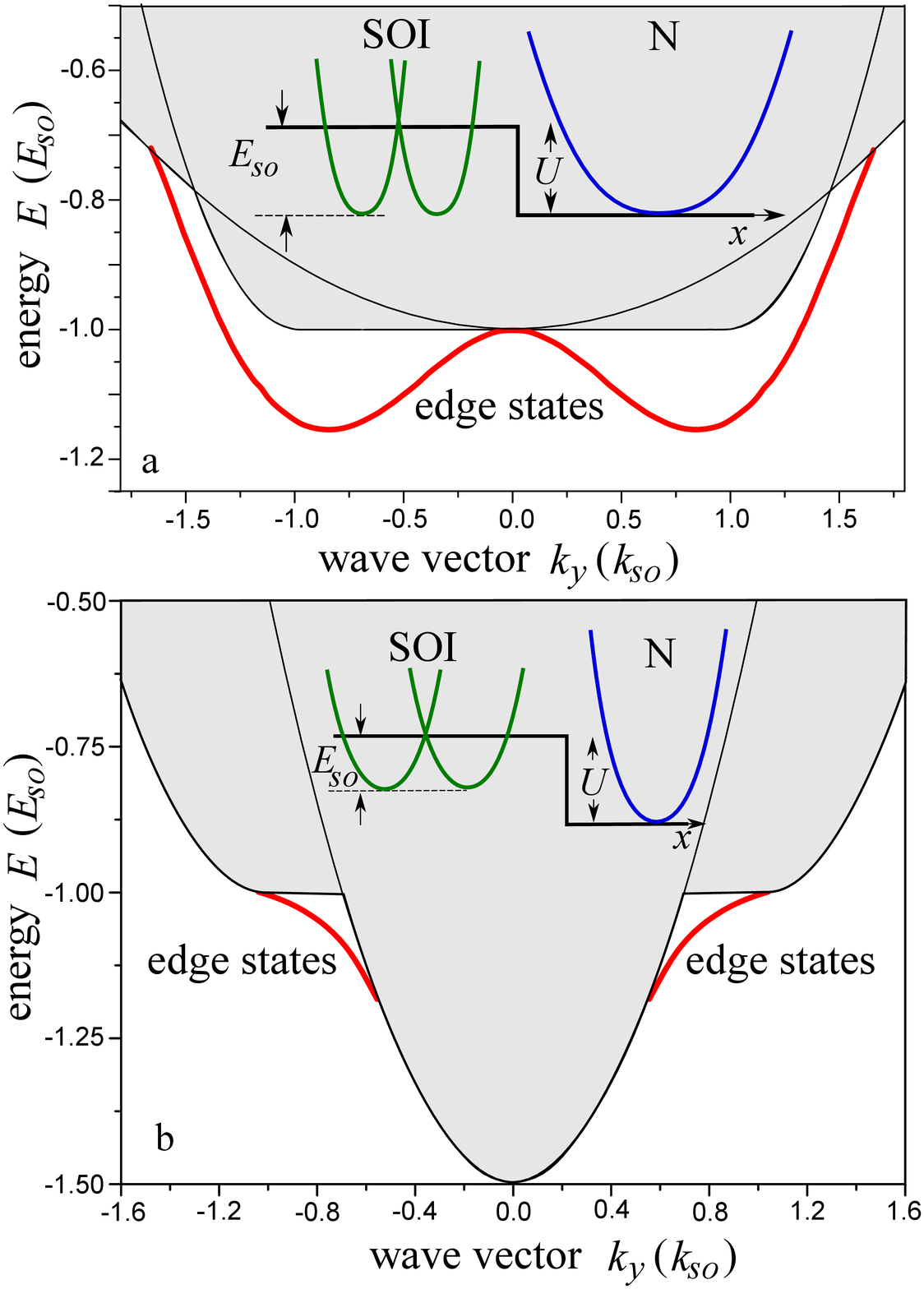}}
\caption{(Color online) Spectra of the interface states at the SOI/N boundary in the conduction band. (a)~Interface-state spectra in the conduction band are continuations of the interface-state spectrum from the forbidden band. $U=E_{so}$ and $\mu=0.1$. (b)~The interface states (two thick line segments) existing only in the conduction band. $U=1.5E_{so}$ and $\mu=1$. Shaded regions indicate the continuum of the bulk states in the SOI and N regions. The insets show schematically the potential shape and the bulk state spectra.}
\label{SOI/N_f2}
\end{figure}

The interface states exist also in the energy region above the conduction band bottom of the bulk states, $E>\mathrm{min}[-E_{so},-U]$. They form here two branches with $k_y<0$ and $k_y>0$. The shape of the dispersion curve in the conduction band depends on the potential step height $U$ and the effective mass ratio $\mu$. There are two kinds of dispersion curves demonstrated in Fig.~\ref{SOI/N_f2}.

Fig.~\ref{SOI/N_f2}a shows the case where the dispersion curve in the conduction band continues the dispersion curve from the forbidden band up to the point of  intersection with the boundary of the bulk-state continuum (Fig.~\ref{SOI/N_f2}a). At this point $\mathrm{Re}\,g=0$.

In Fig.~\ref{SOI/N_f2}b the other case is demonstrated, where the interface states in the forbidden band are absent, but in the conduction band the interface states exist. Their spectrum is presented by two curve segments arranged symmetrically in regions $k_y<0$ and $k_y>0$, as shown in Fig.~\ref{SOI/N_f2}b. These states exist in the energy interval $E_{cr1}<E<E_{cr2}$, with $E_{cr1,2}$ being correspondingly the energies at which the interface-state spectrum intersects the boundaries of the bulk-state continuum in the N- and SOI regions. An equation describing this spectrum is easily found in the case where $\mu=1$
\begin{equation}
\zeta_{ES}=\left(\frac{u-k_{so}^2}{2k_{so}}\right)^2-\frac{k_{so}^{2}k_y^2}{k_{y}^{2}-[(u-k_{so}^2)/2k_{so}]^2}\,,
\label{SOI/N_mu1}
\end{equation}
with additional conditions [Eqs~(\ref{add cond})]. The interface states are absent when these conditions are violated. The interface-state spectrum has two branches corresponding to waves propagating in opposite directions. They are shown in Fig.~\ref{SOI/N_f2}b for the potential step $U=1.5E_{so}$ at the interface. The interface states occupy a finite energy layer and a finite interval of $k_y$. The lower and upper edges of these intervals are determined by the intersection points of the interface-state spectrum with the boundaries of the bulk-state continua in the N and SOI regions.

The lower energy $E_b$ of the interface-state band depends on the potential step $U$ at the interface. The function $E_b(U)$ is easy to find from Eq.~(\ref{eq11}) and the condition $g(\zeta,k_y)=0$:
\begin{equation}
 E_b=-U+\frac{U^2-E_{so}^2}{4E_{so}}.
\end{equation}
It is seen that $E_b=-E_{so}$ at $U=E_{so}$. With increasing $U$, the interface-state bottom $E_b$ decreases to reach the minimum value $E_b=-1.25E_{so}$ at $U=2E_{so}$ and whereupon increases. Thus, the maximum depth of the interface-state bottom is $-0.25E_{so}$ below the conduction band bottom of the SOI region. This conclusion is easy generalized to the arbitrary mass ratio $\mu$.

The interface state formation can be interpreted as a result of the lowering of the electron energy near the interface because of the mutual penetration of electrons from one contacting region to another. Electrons  penetrating from the N region into the SOI region gain energy since they undergo the SOI action. In contrast, the electrons of the SOI region lose energy while penetrating into the N region since they do not feel the SOI there. If $m_{so}\ll m_N$, the electrons penetrate into the SOI region much deeper than into the N region. Hence, the gain in the energy is larger than its loss and a state localized near the interface can appear with energy lower than the conduction band bottom.

\subsection{The spin texture}
The spin texture of the interface states is rigidly connected with the wavevector $k_y$ directed parallel to the boundary.
In the case of the Rashba SOI the spin density vector $\vec{S}(x)$ is directed normally to $k_{y}$, its direction being reversed upon changing the sign of $k_y$. The case of the Dresselhaus SOI is similar, but the the spin vector lies in the plane $(y,z)$.

\begin{figure}
\centerline{\includegraphics[width=0.55\linewidth]{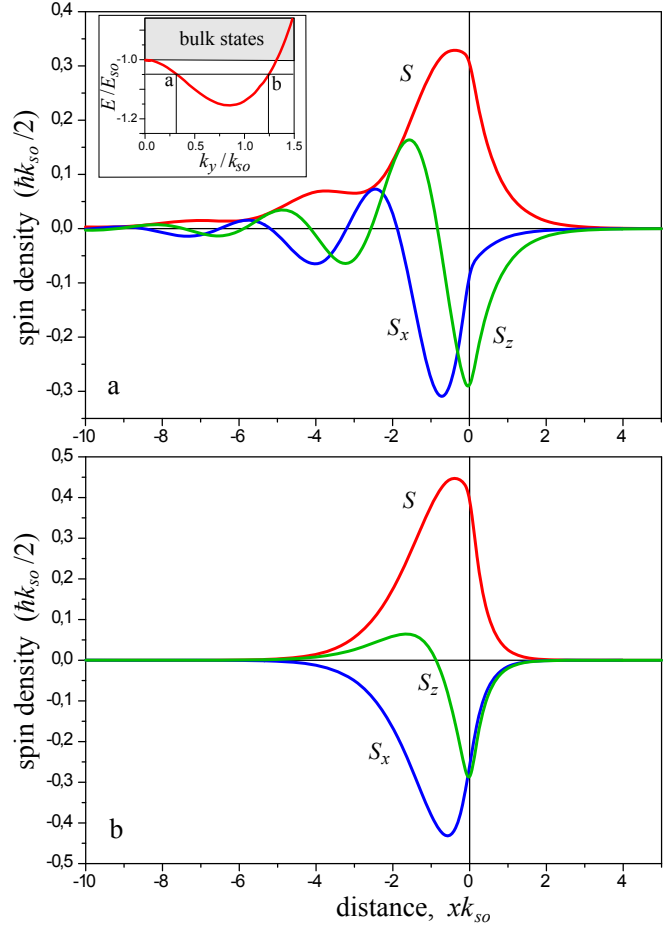}}
\caption{(Color online) Spatial distributions of total spin density $S$ and spin density components $S_x$ and $S_z$ of the interface states with the same energy $\zeta_{ES}=-1.05 k_{so}^2$ and different wavevectors $k_y$: (a) $k_y=0.32 k_{so}$ and (b) $k_y=1.24 k_{so}$. The inset shows the interface-state spectrum and two wavevectors corresponding to the given energy. The parameters used in calculations are $U=E_{so}$, $\mu=0.1$.}
\label{SOI/N_S(x)}
\end{figure}

Below we restrict ourselves by the RSOI case and consider the spatial distribution of the spin density components $S_x$ and $S_z$. Typical dependences of the spin density components $S_{x}(x)$ and $S_{z}(x)$ and the total spin density $S(x)$ on the distance from the boundary are presented in Fig.~\ref{SOI/N_S(x)} for the interface states in the forbidden band whose spectrum is shown in Fig.~\ref{SOI/N_f2}. At a given energy there are two states with different $k_{y}$. They are characterized by a qualitatively different dependence of $\vec{S}$ on $x$ in the RSOI region. In the states with lower $k_y$, the spin components $S_{x}$ and $S_{z}$ oscillate when decaying into the RSOI region. This means that $\vec S$ rotates in the $(x,z)$ plane. In contrast, the spin density in the states with higher $k_y$ decays without oscillations.

Since the spin in the interface states is uniquely connected with the wavevector $k_y$ they transfer a spin current even under the equilibrium conditions, the total spin current of all occupied states being polarized in the $(x,z)$ plane.

\subsection{Interface potential effect}
\label{Interface_potential}

Now turn to effects produced by the interface Hamiltonian~(\ref{boundary_H}). The effect of the spin-independent component of the interface potential $v_0$ consists in shifting the interface-state energy up (when $v_0>0$) or down (when $v_0<0$). Specific calculations carried out in the case of $\mu=0.1$ and $U=E_{so}$ show that (i) the increase in the repulsive potential leads to an increase in the energy of the interface states and finally results in the their disappearance at $\bar{v}_0 =0.7k_{so} $, (ii) the increase in the attractive potential results in shifting the interface-state energy down at such a rate that the energy doubles when $\bar{v}_0=-0.3k_{so}$. This shift of the dispersion curves is accompanied by only a small change in their shape because the potential $v_0$ does not depend on $k_y$.

\begin{figure}
\centerline{\includegraphics[width=0.55\linewidth]{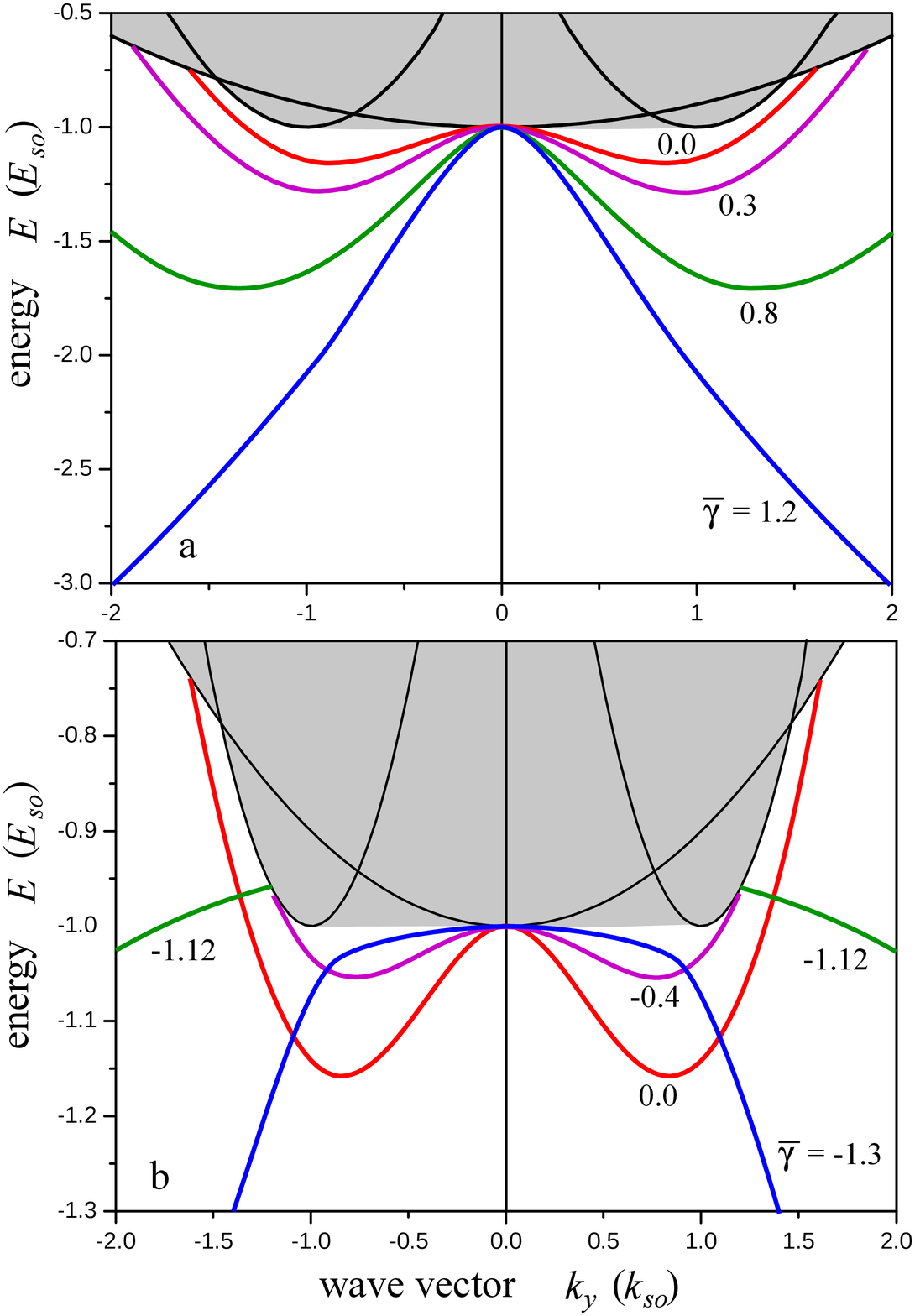}}
\caption{(Color online) Spectrum of the interface states at the SOI/N boundary for a variety of the spin-dependent components of the interface potential: (a) $\bar{\gamma}>0$, (b) $\bar{\gamma}<0$. The parameters used are $\mu=0.1$, $U= E_{so}$, $v_0=0$.}
\label{SOI/N_gamma}
\end{figure}
In contrast, the spin-dependent component of the interface potential [$\gamma \sigma_z k_y$ in Eq.~(\ref{boundary_H})] affects the interface-state spectrum essentially. This effect is demonstrated in Fig.~\ref{SOI/N_gamma} where the interface-state spectra are presented for a variety of $\gamma$. When $\gamma>0$, the increase in $\gamma$ results in lowering the energy of the interface states down to the forbidden band, the energy decrease being stronger for the larger $|k_y|$. Negative $\gamma$ produces a more complicated effect. When $|\bar{\gamma}|\ll 1$, the increase in $|\gamma|$ leads to the growth of the interface-state energies. However, in both cases there is a critical value of $\bar{\gamma}$ above which the energy goes unboundedly to $-\infty$ with increasing $|k_y|$. This means that the states are radically restructured and a many-band consideration is required.

The spin-dependent component of the interface potential also changes the spin texture of the interface states. With increasing $\gamma$ the interface states become more localized near the boundary, the $z$ component of the spin density $S_z$ increases and the component $S_x$ decreases.

\section{Interface states in a RSOI/DSOI contact}
\label{RSOI/DSOI}

Another 2D system in which we demonstrate the existence of the interface states in the forbidden band is a contact of regions with the Rashba and Dresselhaus SOIs (RSOI/DSOI structure). The interface states are studied by solving Eqs.~(\ref{boundary_gen}) and (\ref{wave_func}) in the same manner as described above.

In the Rashba region ($x<0$) the wavefunction is given by Eq.~(\ref{eq2}) with the wavevectors and spin functions defined by Eqs.~(\ref{krx}) and (\ref{chi_R}). The counterparts for the Dresselhaus region ($x>0$) are easily obtained from corresponding expressions for the Rashba region via an unitary transformation ~\cite{Sablikov}.

The electron  wavefunction in the DSOI region is
\begin{equation}
\mbox{\boldmath$\Psi$}^{(D)}=e^{ik_yy}\sum \limits_{s=\pm1}B_{s}\
\left(
\begin{array}{c}
 \chi^{(D)}_s(\vec{k})\\
 1
\end{array}
\right)
e^{\kappa^{(D)}_s x}\,,
\label{eq2D}
\end{equation}
where $\kappa_s^{(D)}=\kappa_1^{(D)}+s\kappa_2^{(D)}$,
\begin{equation}
\kappa_{1,2}^{(D)}\!=\!\frac{1}{\sqrt{2}}\sqrt{k_y^2\!-\!\frac{\zeta\!+u}{\mu}\!-\!2k_{D}^{2}\!\pm \sqrt{\left(\!\frac{\zeta\!+u}{\mu}-\!k_y^2\right)^2\!\!\!-\!4k_{D}^2k_y^2}},
\label{krx2}
\end{equation}
\begin{equation}
 \chi^{(D)}_{s}=\frac{-2ik_{D}\left(k_y+ \kappa_s^{(D)}\right)}{\zeta +u - k_y^2+ \left.\kappa_s^{(D)}\right.^2}\,,
\end{equation}
$k_{D}=m_D\beta/\hbar^2$,  $\mu=m_{R}/m_D$, and $m_R$ and $m_D$ are the effective masses of electrons in the RSOI and DSOI regions.

\begin{figure}
\centerline{\includegraphics[width=0.55\linewidth]{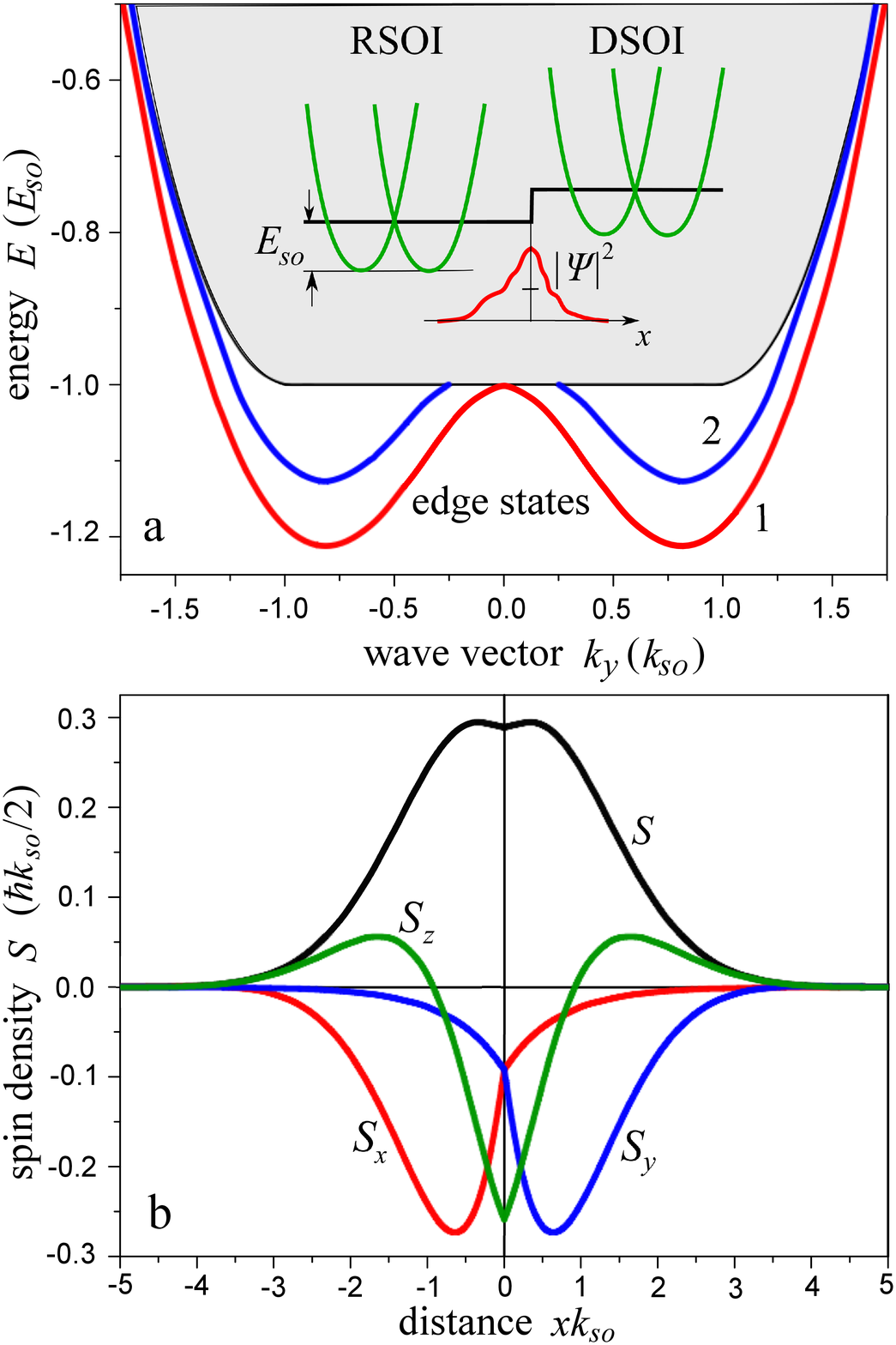}}
\caption{(Color online) (a) Spectrum of the interface states in the RSOI/DSOI structure for two potential steps: $U_D-U_R=0$ (line 1) and $U_D-U_R=0.25E_{so}$ (line 2). Shaded regions indicate the bulk states. Inset: the potential shape, the bulk-state spectra and the electron density distribution in the interface state. (b) Spatial distribution of the spin density components ($S_x, S_y, S_z$) and the total spin density $S$ in the interface state with the energy $\zeta_{ES}=-1.2k_{so}^2$ and momentum $k_y=0.95k_{so}$. The parameters used are $U_D=U_R$, $\alpha=\beta$, $m_R=m_D$.}
\label{RSOI/DSOI_f}
\end{figure}

The interface-state spectrum is calculated ignoring the interface potential. The results are presented in Fig.~\ref{RSOI/DSOI_f} for two different potential steps at the boundary.

The interface states are seen to exist in the forbidden band even if the effective masses in the contacting regions are equal, in contrast to the case of the SOI/N system. But the energy interval, where the interface states are located, and a general view of the spectra are quite similar to those shown in Fig.~\ref{SOI/N_f2} for the SOI/N structure. The origin of the interface states can be interpreted as a result of the mutual action of the SOIs in the two contacting regions.

The electron and spin densities in the interface states are localized near the boundary at a distance of the order of the characteristic SOI length. It is worth noting that the $S_x$ component of the spin density is concentrated in the RSOI region whereas the $S_y$ component is located mainly in the DSOI region (Fig.~\ref{RSOI/DSOI_f}b). The spin direction is reversed when the sign of $k_y$ is changed. Therefore the interface states are of chiral nature. In addition, the interface states carry a spin current under thermal equilibrium. Under the nonequilibrium conditions appearing when a particle current flows parallel to the boundary, the spin density accumulates in the interface states.

\section{Quantization of the interface states in a strip structure, DSOI/RSOI/DSOI}
\label{Strip}

In a strip of electron gas with SOI the interface states exist near both opposed boundaries. In this section we study quantum states in the case where the strip width is of the order of their localization length. Under such conditions two effects are important: the size quantization of the interface states and the overlap between the states located near the opposite sides.

We have considered the strip structures of two types: the strip of the 2D electron gas with the Rashba SOI bounded laterally by regions with the Dresselhaus SOI (DSOI/RSOI/DSOI structure) and the RSOI strip bounded by the normal electron gas (N/RSOI/N structure). In the case of strip structures the calculations are more cumbersome than above since the wavefunctions are to be found in three regions, using two boundary conditions of the form of Eq.~\ref{boundary_gen}. We have solved these problems numerically and present below the results only for the DSOI/RSOI/DSOI structure. In the case of the N/RSOI/N structure the results are similar.

\begin{figure}
\centerline{\includegraphics[width=0.55\linewidth]{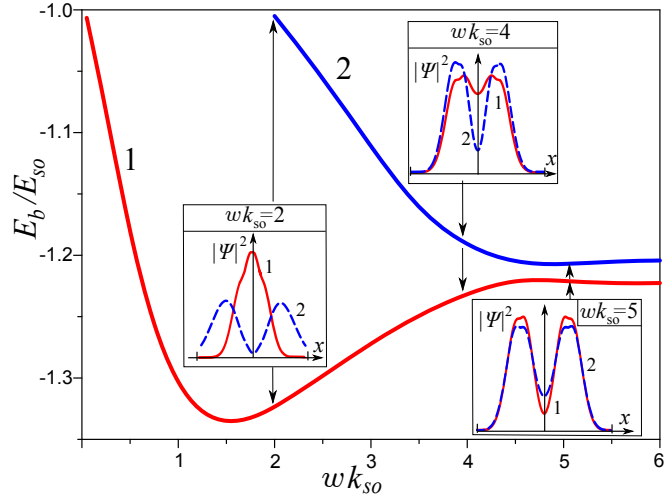}}
\caption{(Color online) Energy of the interface-state subband bottoms as functions of the strip width $w$ in the DSOI/RSOI/DSOI. In the insets, the electron density distribution in the interface states is shown for different widths of the strip: $wk_{so}=2, 4, 5$. Lines 1 and 2 (dashed) correspond to the lower and upper subbands. The parameters used are $U_D=U_R$, $\alpha=\beta$, $m_R=m_D$.}
\label{DRDwidth}
\end{figure}

The main difference between the spectra of the interface states in the strip structures and in the single RSOI/DSOI contact, lies in the splitting of the interface-state band into two subbands with different distributions of the electron density across the strip. This is demonstrated in Fig.~\ref{DRDwidth} where the energies $E_{b1}$ and $E_{b2}$ of the interface-state band bottoms are drawn as functions of the strip width $w$ in the case where $\alpha=\beta$, $m_R=m_D$, $U_D=U_R$. In this case the SOI wavevectors in all regions are equal $k_R=k_D=k_{so}$.

It is interesting that the quantization energy depends on the width in an unusual manner. The lower subband bottom $E_{b1}$ decreases with increasing $w$ until $wk_{so}\lesssim 1.5$. This is trivially explained by the decrease of the kinetic energy. However, $E_{b1}$ unexpectedly grows as $wk_{so}>1.5$.

To interpret such a behavior of the quantized energy let us take into consideration the fact that the electron density is redistributed across the strip with increasing $w$ as shown the insets in Fig.~\ref{DRDwidth}. The electron density in the lower subband is redistributed from the center to the edges. As it has been discussed above, electrons gain energy near the interface because of the mutual action of the SOIs in the two regions. In the case of the strip, there is an additional energy gain caused by the joint effect of two interfaces. This energy gain decreases with increasing $w$ because the interface states overlap less. It is for that reason $E_{b1}$ grows with $w$. In the case of the upper subband this effect is much smaller since the electron density in these states is always concentrated closer to the edges.

It is worth noting that this effect results in essential lowering of the interface-state energy down to the forbidden band as compared with the case of a single interface. This unusual behavior of the quantization energy can manifest itself in the formation of localized states in quantum constrictions whose width varies slowly with the longitudinal distance.

\section{Concluding remarks}

In this paper we have shown that the interface states exist at a boundary between regions with different SOIs in 2D electron systems and studied their spectra and spin texture. Depending on the system parameters, the energy of these states can lie either in forbidden or the conduction band, or in both bands. An unexpected result is that the interface states arise even if the effective potential of the interface equals zero. The interface states are similar to the edge states in 2D topological insulators since they have chiral spin texture and are determined by the bulk properties of materials, such as the SOI strengths and the effective mass ratio.

In view of the interest to the robustness of the topological edge states, the persistence of the interface states found here within an one-band model for variation of the boundary conditions has been explored.

We have studied the effect of the interfacial potential, which has two components: spin-dependent and spin-independent ones. The latter component does not qualitatively affect the interface-state spectrum, while the spin-dependent component changes the states essentially as this potential exceeds a critical value.

The persistence of the interface states for smoothing of  the boundary was also investigated. We studied the system in which the SOI strength changes  smoothly in the transition layer between the regions with different SOIs and found that the interface states survive if the transition layer width $L$ is small compared to the characteristic length of the SOI, $Lk_{so}<1$. With increasing layer width the energy interval where the interface states are located diminishes and finally the states disappear.

Thus, we conclude that the interface states in the system with heterogeneous SOI are less robust than the edge states in 2D topological insulators. In addition, their spectrum contains four states in the forbidden band at a given energy and hence there are more possibilities for scattering.

In realistic systems the SOIs of the Rashba and Dresselhaus types often act simultaneously and therefore it is expedient to explore this situation. When both types of the SOI are present the electron spectrum becomes anisotropic in the wavevector space, giving rise to interesting effects.~\cite{Averkiev} Particularly, in the case where $\alpha=\beta$, the SOI is effectively suppressed for electrons moving along the [110] direction in a zinc-blende crystal. The detailed study of the interface states in 2D systems with both SOIs is beyond the scope of the present publication. Here we restrict ourselves by a brief discussion of results obtained for the contact of the (R+D)SOI region with normal electron gas. The joint action of the Rashba and Dresselhaus SOIs turns out to  always reduce the energy depth of the interface states below the bottom of the bulk continuum states, the effects of the Rashba and Dresselhaus SOIs being not completely balanced at any ratio of $\alpha$ and $\beta$. The interface state energy depends on the orientation of the interface line. The maximum energy depth of the interface states is reached when the boundary is parallel to the [1\=10] direction. For any orientation of the boundary, there is such a ratio of $\alpha$ to $\beta$ at which the interface state energy intersects the conduction band bottom. In the particular case where $\alpha = \beta$, below the conduction band bottom the interface states are absent.

There is another effect close to that considered in this paper. It consists in the existence of the edge spin currents in 2D systems with heterogeneous SOI. The mechanism of the edge spin currents  is caused by the spin-dependent scattering of bulk electrons on the interface ~\cite{Sablikov1}. These currents also exist under the equilibrium. The estimations show that the spin current of the interface states studied here is small compared to scattering spin current if the Fermi energy lies higher than $E_{so}$ above the conduction band bottom. In opposite case, the interface-state spin current is prevailing, especially when the Fermi level lies in the forbidden band.

\subsection*{Acknowledgments}
This work was supported by Russian Foundation for Basic Research (project No~11-02-00337) and Russian Academy of Sciences (programs ``Basic researches in nanotechnology and nanomaterials'' and ``Strongly correlated electrons in solids and structures'').

\end{document}